# Title：Exploring the determinants on massive open online courses continuance learning intention in business toward accounting context

**Author's name:** D. SHANG*[a,b], Q. CHEN[a], X. GUO [a], H. JIN[a], S. KE [a], M. LI [a]
**Affiliation:**
[a] Business School, Zhengzhou University, Zhengzhou
[b]
[b] GTSI & College of Computing, Georgia Institute of Technology
**Postal address:**
[a] Zhengzhou University, Business School, Zhengzhou 450001, Peoples R China.
[b] GTSI & College of Computing, Georgia Institute of Technology.
**Corresponding author email address**: shangdw@zzu.edu.cn&13895787539@126.com (D. SHANG)

**Acknowledgement:**
XX……

**Abstract:** Massive open online courses (MOOC) have become important in the learning journey of college students and have been extensively implemented in higher education. However, there are few studies that investigated the willingness to continue using Massive open online courses (MOOC) in the field of business in higher education. Therefore, this paper proposes a comprehensive theoretical research framework based on the Theory of Planned Behavior (TPB). In the field of business, a representative accounting course is taken as an example. We adopt the questionnaire survey method and use the partial least squares structural equation model to analyze the collected feedback data from college students and test the hypotheses. This paper focuses on the potential influencing factors and mechanisms of the willingness to continuously use Massive open online courses (MOOC) in accounting. The results show that interface convenience (IC) and interface design aesthetics (IDA) have positive effects on user attitude (ATT). User attitude (ATT), perceived behavioral control (PBC), and subjective norms (SN) have positive effects on the continuance learning intention. In addition, academic self-efficacy (EF) not only significantly affects continuance learning intention (CI) but also moderates the relationship between the Theory of Planned Behavior (user attitude, perceived behavior control, subjective norms) and the continuance learning intention of accounting MOOC. Therefore, the Theory of Planned Behavior(TPB) is extended in social science accounting Massive open online courses environment. Based on these findings, this paper provides several theoretical and practical implications for researchers and practitioners of MOOC, accounting, and the design of learning systems in higher education contexts.

**Keywords:** higher education; the Theory of Planned Behavior; Massive open online courses (MOOC); academic self-efficacy; continuance learning intention

## 1. Introduction

As ubiquitous computing and information technologies develop rapidly, Massive open online courses (MOOC) have become an important part of the learning process and have been broadly implemented by higher education institutions (Tzeng et al., 2022; Wu & Chen, 2017). MOOC have attracted attention, which has drove the advancement of remote learning. (Ma & Lee, 2019; Pozon-Lopez et al., 2020). As we navigate through this new era, MOOC have emerged as a prominent tool in education, offering flexible and accessible learning opportunities for learners worldwide (Chen et al., 2020; Sue G, 2022). Research on Massive open online courses (MOOC) started in 2008 and gained significant popularity by 2016, causing a potential threatening to the global educational economy (Liu et al., 2022; Tatnall Arthur et al., 2022). This popularity can be attributed to the fact that MOOC are combined with a wide range of social networks, resulting a major impact：on education. These courses offer learners rich resources, including instructional videos, learning content and additional material. Moreover, they establish discussion boards and take advantage of multiple digital networks and online learning platforms to promote interaction and collaboration among instructors, learners, and the course itself. Many educational institutions incorporate MOOC into their curriculum, which effectively organize and establish connections with a large group of online learners pursuing similar targets. Because of the impact of MOOC' low course completion rates and frequent breaks on learning outcomes, researchers in digital learning and related disciplines have simultaneously focused on students' intentions to continue learning (CI) (Psotka & Chen, 2019; Tzeng et al., 2022). Learners in MOOC have less behavioral perseverance. To increase course completion rates and continuity, it is essential to investigate the factors affecting learners' inclinations to keep studying (Jung & Lee, 2018). By understanding the influencing factors such as learning motivation, learning environment design, and individual behavioral characteristics, it is possible to provide theoretical guidance for the improvement and optimization of MOOC learning environments. This further stimulates students' enthusiasm for learning and enhances their intention to continue using MOOC.

Previous empirical studies on the intention to continue using MOOC have primarily focused on two aspects: learner characteristics such as perceived usefulness, satisfaction, and self-efficacy (Jin & Shang, 2022; Rekha et al., 2023), and technical characteristics such as task technology fit, personalized design (Shin & Song, 2022; Li et al., 2022; Rohan et al., 2021). Studies have demonstrated that learners' personal internal and external factors have positive effects, either directly or indirectly, on their willingness to continue learning in MOOC (Jung & Lee, 2018; Psotka & Chen, 2019). However, there is limited research specifically examining influencing factors in different disciplines. Most studies focus on science, engineering, and linguistics courses (Apricio et al., 2019; Li et al., 2018; Psotka & Chen, 2019). As a result, further exploration is needed in business subjects such as accounting. Taking accounting as an example, accounting MOOC typically serve as core courses for accounting majors, with the goal of imparting fundamental financial and accounting knowledge and improving students' understanding of accounting language and business terminology. When it comes to business education, the goal of accounting MOOC courses is to improve students' business literacy and overall capabilities, while accounting itself is unique in terms of content, professional requirements, and practical applications. Nevertheless, business MOOC, especially accounting courses, is different significantly from skill-based science and engineering courses regarding learning motivation and course design. Previous studies did not consider potential influencing factors, such as proprietary business languages like accounting entries and learners' learning efficacy. Thus, this study raises two essential questions: "What factors influence the willingness to continue learning in accounting MOOC for business?" and "What is the link between these possible influencing factors and intention to continue learning?"

This study aims to discover the factors that affect the intention to keep using MOOC in business accounting education and to understand the underlying mechanisms. Figure 1 presents an integrated theoretical research framework based on the Theory of Planned Behavior (TPB) and the system implementation perspective (Hsiao, 2013; Wu & Shang, 2019). The study employed questionnaires and a PLS-SEM (Partial Least Squares Structural Equation Modeling) to analyze responses from 176 college students at a provincial university in China. By discussing the possible influencing factors and their relationships, this paper endeavors to improve the user experience and enhance the willingness to continue using MOOC. The subsequent sections of this article are organized in the following order: section 2 provides a detailed explanation of the theoretical foundation of the research framework and the hypotheses tested. Section 3 describes the study methods employed. Section 4 evaluates the obtained results. Finally, section 5 summarizes the contributions, restrictions, and future prospects of this study.

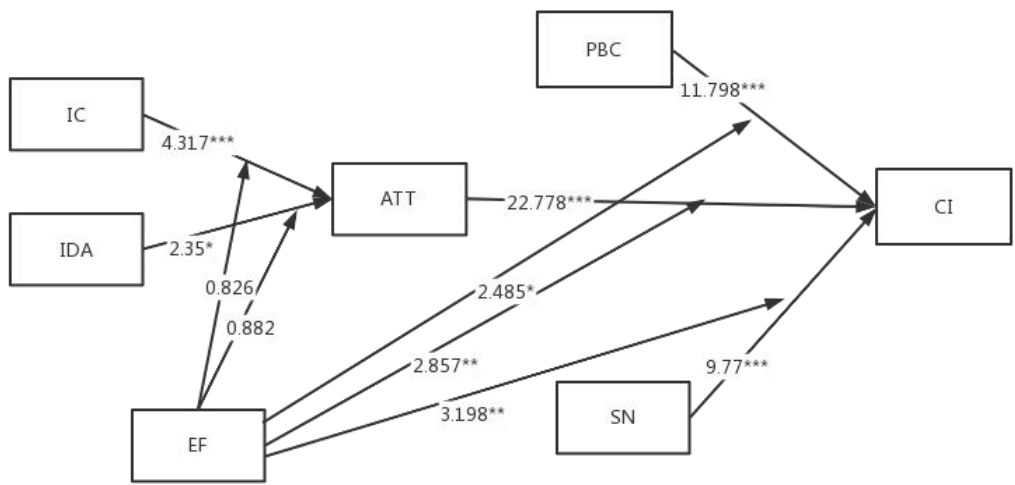

**Figure 1. Research model**
(Note:IC,Interface convenience; IDA,Interface design aesthetics; ATT,attitude; PBC,Perceived behavioral control; SN,Subjective norm; EF, Academic self-efficacy; CI,continuance learning intention)

## 2. Theoretical framework and hypothesis formulation
### 2.1 Accounting MOOC education and Continuance learning intention

As a flexible and convenient form of online learning, MOOC offer students limitless access to high-quality educational resources worldwide and have been successfully applied across various disciplines (Robinson, 2016). However, they also face challenges such as low participation rates, low completion rates, and high dropout rates (Gu et al., 2021; Daneji et al., 2019). Nonetheless, scholars widely acknowledge that learners' willingness to continue using MOOC is a critical factor (Rekha et al., 2023; Kim and Song, 2022). Continuance learning intention (CI) refers to the willingness of learners to persist in using and engaging with MOOC after their first exposure (Al-Mekhlafi et al., 2022). Understanding learners' usage intentions is crucial for educators to optimize instructional design and enhance motivation and learning outcomes (Shanshan & Wenfei, 2022; Yang & Lee). As a result, it is essential to investigate the aspects affecting learners' intentions to continue using MOOC. Previous studies have identified perceived usefulness, satisfaction, self-efficacy, task-technology fit, personalized design, among others, as factors influencing learners' continued usage intentions (Jin & Shang, 2022; Rekha et al., 2023; Shin & Song, 2022; Li et al., 2022; Rohan et al., 2021).

However, existing research has several limitations. Firstly, there is a lack of specific investigation into MOOC in the business field, particularly accounting. Compared to other disciplines, accounting possesses unique characteristics. Thereby it requires careful analysis when applying findings from other areas. Secondly, current research primarily focuses on the influence of individual learner characteristics, educational system characteristics, and environmental factors on continued usage intentions. The understanding of the interaction and complex relationships among these factors is not yet thorough enough. Hence, further studies are needed to explore the formation mechanisms and influencing factors of continued usage intentions in MOOC.

### 2.2 Hypothetical development

As the rapid progress of information technology, MOOC have emerged as a significant mode of contemporary education. In the field of accounting discipline, the application scope of MOOC is gradually expanding. It should be noted that learners' attitudes play a significant role in the effective application of MOOC. Moreover, research has shown that interface convenience (IC) is a significant factor influencing learners' attitudes towards MOOC and their willingness to continue using. Interface convenience (IC) is also a key concept to consider when designing and building user interfaces. Its purpose is to enhance users' interaction with computer systems, software applications, or other digital products by providing them with a user-friendly operating experience. It can reduce cognitive load and operational difficulties, as well as improve user satisfaction and efficiency.

When users utilize the accounting MOOC learning system, interface convenience allows them to easily locate the necessary information and quickly find the required learning resources, course content, and task requirements. Additionally, users can interact with lecturers and seek guidance through message boards and other means, enabling timely support and fostering interactive learning. In this context, we propose Hypothesis 1:

H1: Interface convenience has a positive impact on user attitude.

By incorporating various elements, such as color settings and balanced layouts, well-designed interfaces can provide a pleasant user experience and promote users' emotional and cognitive levels. These aesthetic elements not only make the interface visually appealing but also have an impact on users' emotions and psychological states. Different colors cause different emotional reactions to individuals. Proper colour schemes can be effective in regulating the mood of the user, which in turn fosters a positive attitude to MOOC. What's more, a balanced layout design can contribute to a clean and organized interface, preventing user confusion. The clarity of the layout improves the understanding of the user, fosters confidence in the MOOC, and improves satisfaction with the overall user experience. Hence, we propose Hypothesis 2:

H2: Interface design aesthetics has a positive impact on user attitude.

The Theory of Planned Behavior, put forward by psychologist Icek Ajzen (2020), is a theoretical model that explains individual behavior. It has a close relationship with the intention to continue using accounting MOOC. Based on this theory, person's behavior is influenced by three main factors: attitude, perceived behavioral control, and subjective norms. By applying the Theory of Planned Behavior, we can gain deeper insights into the effects of attitude, perceived behavioral control, and subjective norms on the intention to continue using accounting MOOC. This allows us to effectively predict and explain user behavior.

In terms of user attitudes, the Theory of Planned Behavior indicates that attitude is key to behavior intent and decision making. In accounting MOOC, the attitude of an individual relates to the perception, opinion, and subjective assessment of the platform. Knowing how users feel about accounting MOOC can help determine their preferences, which will guide the design and improvement of the platform to increase users' satisfaction. If MOOC interface is thought as convenient and aesthetically pleasing, with high quality teaching content and strong interaction, the users might develop a positive attitude (Roos and Hahn, 2019). This positive attitude is critical to their intention to continue using these platform. When individuals possess a favorable attitude towards accounting MOOC, they have more possibilities to maintain motivation for continued use because they believe the platform can meet their learning needs and help improve their accounting knowledge (Lee, 2010).

Perceived behavioral control (PBC) refers to individuals' beliefs and abilities to perform specific behaviors. When studying whether to continue to use MOOC, it is important to pay attention to the user's perception of their skills, the availability of learning resources, and the challenges and barriers they face. It is essential to know how users perceive those factors, which can help identify their needs and problems and allow for targeted support and solutions to promote their continued use of accounting MOOC. If individuals believe they have the necessary skills, resources, and control to continue using accounting MOOC, they are more likely to exhibit more positive intentions to continue. Meanwhile, when individuals are confident and believe they can overcome difficulties, they are more likely to persist in using accounting MOOC and achieve the expected learning outcomes. Subjective norm (SN) refers to individuals' perception of others' expectations and pressure regarding a specific behavior in the social environment. Identifying possible social and external impacts may inform the introduction of appropriate policies for the promotion of MOOC, which in turn will enhance the purpose of continuous learning. In the context of the intention to continue using accounting MOOC, subjective norms may include recommendations from others, peers' usage experiences, and advice from professional instructors. For example, when the course instructor supports their use of MOOC for learning, users are more likely to maintain continuance learning intention. Based on the above mentioned theoretical frameworks, these following hypotheses are proposed:

H3: Users' attitudes towards accounting MOOC contribute positively to their intention to continue using them.

H4: Perceived behavioral control has a positive influence on users' continuous using intention towards accounting MOOC.

H5: Subjective norms play a positive impact on users' intention to continue using accounting MOOC.

Academic self-efficacy is defined as individuals' belief in their own competence and ability to accomplish specific tasks. It is a subjective perspective that reflects beliefs and expectations of whether one can successfully achieve goals in the learning process. It is closely related to individuals' positive attitudes towards learning. When individuals own high academic self-efficacy, they have more confidence in their ability to effectively complete learning tasks. A learning platform with good interface usability provides users with a user-friendly interface, a clear layout, and easy navigation, making it easier for individuals to use the platform for learning. Therefore, students with high academic self-efficacy will be satisfied and enjoy the learning process when they use the platform which has a high level of usability (Lee, 2010). So academic self-efficacy moderates the influence of interface usability on the continuous willingness to use accounting MOOC. Specifically, higher academic self-efficacy can enhance individuals' sensitivity to interface usability, making them more actively accept and utilize the learning experience provided by good interface usability. This further increases their intention to continue using accounting MOOC.

Furthermore, academic self-efficacy may moderate the impact of interface design aesthetics on the continued intention to use accounting MOOC by providing cognitive abilities, increasing learning motivation, and promoting a good user experience. When people with high academic self-efficacy use accounting MOOC and believe that they can master accounting knowledge and skills effectively, they have more possibilities to have a positive perception of interface design aesthetics and continue using the MOOC. Moreover, a well-designed interface can provide a clear and intuitive functional layout and operation process, making the learning process smoother and more enjoyable. Users with a beneficial experience are more likely to have a positive assessment of the aesthetics of the interface and to continue to use the MOOC. Therefore, the following hypotheses are proposed:

H6a: Academic self-efficacy positively moderates the influence of interface usability on the sustained motivation to use accounting MOOC.

H6b: Academic self-efficacy positively moderates the impact of interface design aesthetics on the continued intention to use accounting MOOC.

The Self-Determination Theory proposes that when users' psychological needs for autonomy, relatedness, and competence are satisfied, they will be motivated to participate more actively and continuously. In the area of education and learning, understanding learners' intrinsic motivation and meeting their needs for autonomy, relatedness, and competence can help designers create a positive learning environment, stimulate learners' initiative and interest, and improve learning outcomes. Therefore, when learners have high academic self-efficacy, they will be more confident in their abilities, show a more positive learning attitude, and choose to continue using accounting MOOC platforms (Khan et al., 2023). At the same time, academic self-efficacy encourages users to participate more actively in the learning process, investing more time and experiences, leading to more active participation in the learning process. Academic self-efficacy can also establish a virtuous cycle. When learners achieve certain learning outcomes in accounting MOOC, their academic self-efficacy will further increase, providing more motivation to continue using the accounting MOOC platform. In terms of particular assignments, academic self-efficacy is described as individuals' confidence and beliefs in their ability to succeed. Perceived behavioral control (PBC) refers to individuals' perceptions of their ability to control and influence specific behavioral outcomes. There is a certain influence between these two factors. When learners achieve good results, understand difficult points, and feel their abilities have improved in accounting MOOC, their academic self-efficacy is strengthened. Learners believe in their ability to master and apply accounting knowledge, which enhances their perceived behavioral control. When learners have higher perceived behavioral control, they believe that they can effectively control their learning situation and are more likely to maintain their intention to continue using accounting MOOC. They believe that their behavior can generate positive results, and therefore, they are inclined to continue using the platform to obtain more learning achievements. As a result, by enhancing academic self-efficacy, perceived behavioral control can be moderated, thereby positively influencing individuals' continued intention to use accounting MOOC (Maqableh et al., 2021).

Previous research has shown that learners' academic self-efficacy is enhanced when they achieve good learning results in accounting MOOC. Academic self-efficacy influences users' subjective norms towards continuous use of accounting MOOC (Ni and Cheung, 2023). When learners is convinced they can understand accounting knowledge, apply what they have learned, and perceive these behaviors as meeting their learning needs and goals, their subjective norms tend to prefer continuing use. When learners' subjective norms align with the intention to continue using accounting MOOC, they are more likely to maintain their willingness to continue using them because it aligns with their expectations of self-ability and goals. Therefore, the following hypotheses are proposed:

H6c: Perceived academic self-efficacy positively moderates the influence of user attitude on the willingness to continue using accounting MOOC.

H6d: Perceived academic self-efficacy positively moderates the influence of perceived behavioral control on the willingness to continue using accounting MOOC.

H6e: Perceived academic self-efficacy positively moderates the influence of subjective norms on the willingness to continue using accounting MOOC.

To strengthen the consistency between the hypotheses and acronyms, a hypothesis and vocabulary list of all abbreviations mentioned in this study is summarized in Table 1.

Table 1. Hypotheses and acronyms

| Hypotheses and relationship | | Full term and acronyms | |
| --- | --- | --- | --- |
| Hypothesis | Relationship | Full term | Acronyms |
| H1 | Interface convenience has a positive impact on user attitude | Interface convenience | IC |
| H2 | Interface design aesthetics has a positive impact on user attitude | Interface design aesthetics | IDA |
| H3 | Users' attitudes towards accounting MOOC contribute positively to their intention to continue using them | Attitude | ATT |
| H4 | Perceived behavioral control has a positive influence on users' continuous using intention towards accounting MOOC | Perceived behavioral control | PBC |
| H5 | Subjective norms play a positive impact on users' intention to continue using accounting MOOC | Subjective norm | SN |
| H6a-e | Academic self-efficacy positively moderates the influence of interface usability on the sustained motivation to use accounting MOOC | Academic self-efficacy | EF |

D. SHANG*a,b, Q. CHENa, X. GUO a, H. JINa, S. KE a, M. LI a

## 3. Methodology

This study employed a questionnaire survey as the data collection method and utilized eight modified and adapted constructs as measurement scales (Cho et al., 2009; Deitz et al., 2018; Huang, 2012; Lee, 2010; Shang & Wu, 2017; Tavani et al., 2016). The measurement scales used a 5-point Likert scale, where "1 = strongly disagree" and "5 = strongly agree,". After the development of questionnaire tool, it underwent evaluation and assessment by five scholars who have extensive experience in the field of MOOC. Based on their feedback, modifications were made to the questions and descriptions to enhance comprehension. Before formally completing the questionnaire, respondents were asked if they frequently participated in accounting MOOC or if they wished to register and immediately experience accounting MOOC through a recommended Uniform Resource Locator (URL).

This study utilized a two-stage cluster sampling method to select university students from a provincial university in China as the sample population (Kim et al., 2021). At the first step, 70 university students were selected using cluster sampling from a provincial university in China. We obtained permission from the IT collaboration providers of these universities to access student email addresses. The preliminary survey was conducted with the 70 university students. The results indicated that all measurement scales met the relevant standards for carrying out a wide-ranging survey. At the second step, based on the sample from the first step, questionnaires were distributed to the participants using simple random sampling with the help of database computation techniques (Kim et al., 2021). A total of 177 questionnaires were distributed, with 176 valid responses received, completing the two-step sampling survey accordingly. By employing two-step sampling and assessing the quality of the sampling service, (Kim et al., 2021), sampling errors were successfully minimized, and the satisfied sampling results were obtained. The questionnaire survey spanned a period of approximately four weeks, from May 1, 2023, to May 30, 2023. Empirical data were collected from the respective participants.

## 4. Results
### 4.1 Measurement model

Partial Least Squares (PLS) is a widely accepted and applied non-parametric statistical method that is suitable for research in the social sciences (Chen et al., 2014; Chen et al., 2012; Dijkstra & Henseler, 2015; Hair et al., 2016). Therefore, in this study, PLS-SEM and the SmartPLS 3.0 software were chosen for analysis to evaluate learners' intentions to continue using the survey. The results of the reliability results on the latent variables are presented in Table 2a. All the latent variables exhibit Cronbach's alpha coefficients that are equal to or better than 0.7, which indicates good reliability. This study employed composite validity and discriminant validity to test the validity of latent variables. The values, both average variance extracted (AVE) and composite reliability, surpass 0.5 and 0.7 respectively (Fornell & Larcker, 1981). Table 2b further demonstrates the dependability and accuracy of the latent variables. Discriminant validity is tested according to the Fornell-Larcker criterion (Hair et al., 2016). Therefore, the square root of average variance extracted for each construct is superior to the correlation with other constructs, confirming the effectiveness of discriminant validity (Fornell & Larcker, 1981). All criteria are satisfied, therefore, Reliability results tests and discriminant validity tests are reliable.

**Table 2a. Reliability test results**

| Construct | Cronbach's alpha | Composite reliability | Average variance extracted |
|---|---|---|---|
| ATT | 0.984 | 0.989 | 0.969 |
| CI | 0.951 | 0.969 | 0.912 |
| EF | 0.976 | 0.981 | 0.911 |
| IC | 0.967 | 0.979 | 0.939 |
| IDA | 0.944 | 0.964 | 0.899 |
| PBC | 0.965 | 0.977 | 0.935 |
| SN | 0.968 | 0.979 | 0.939 |

**Table 2b. Discriminant validity (Fornell-Larcker Criterion)**

|  | ATT | CI | EF | IC | IDA | PB | SN |
|---|---|---|---|---|---|---|---|
| ATT | 0.984 | | | | | | |
| CI | 0.948 | 0.955 | | | | | |
| EF | 0.825 | 0.808 | 0.955 | | | | |
| IC | 0.882 | 0.877 | 0.877 | 0.969 | | | |
| IDA | 0.858 | 0.854 | 0.859 | 0.934 | 0.948 | | |
| PBC | 0.924 | 0.910 | 0.858 | 0.897 | 0.874 | 0.967 | |

| | SN | 0.912 | 0.877 | 0.842 | 0.841 | 0.849 | 0.909 | 0.969 |

*4.2 Structural model*

By performing a structural model analysis of the relationships between various latent variables and evaluating the importance of path coefficients, we tested the hypotheses. It was done using a resampling method based on bootstrapping with 176 resamples to obtain stable estimates (Hair et al., 2016). Accordingly, Figure 2 presents the t-values and importance of the path coefficients. The results showed that all direct effects were significant, and all moderation effects were significant except for H6a and H6b, providing support for most of the hypotheses. The path coefficient with the highest t-value in the direct effects was 22.778, reflecting the relationship between perceived behavioral control and continuance learning intention. The path coefficient with the lowest t-value in the direct effects was 2.35, reflecting the connection between interface aesthetics and continuance learning intention. The path coefficient with the greatest t-value in the moderation effects was 3.198, indicating the important moderating impact of academic self-efficacy on the relationship between subjective norms and intention to continue using accounting MOOC. The path coefficient with the lowest t-value in the moderation effects was 0.826, indicating that academic self-efficacy does not significantly moderate the relationship between interface convenience and intention to continue using accounting MOOC.

Interface convenience and interface design aesthetics significantly influence intention to continue using accounting MOOC ($p < 0.001$; $p < 0.05$), supporting H1 and H2. User attitudes, perceived behavioral control, and subjective norms all exert a substantial influence on the intention to persist in using accounting MOOC ($p < 0.001$; $p < 0.001$; $p < 0.001$), supporting H3, H4, H5. These are important influencing factors for intention to continue using accounting MOOC, as they exert influence on individuals' decision-making and behavior in different ways, thereby affecting intention to continue using accounting MOOC. Built on the Theory of Planned Behavior, academic self-efficacy significantly moderates the impacts of user attitudes, perceived behavioral control, and subjective norms on continuous intention to use accounting MOOC ($p < 0.01$; $p < 0.05$; $p < 0.01$), supporting hypotheses H6c, H6d, H6e. Hence, academic self-efficacy, which acts as a moderator of the effects of user attitudes, perceived behavioral control, and subjective norms on intention to continue using accounting MOOC, reflects people's confidence in their capacity for learning and the significant influence of their assessment on intention.

Unexpectedly, academic self-efficacy doesn't significantly moderate the relationship between interface usability and interface design aesthetics with the intention to continue using MOOC ($p > 0.05$), rejecting H6a and H6b. This insignificant finding can be analyzed within the context of MOOC learners' experiences and backgrounds. On the one hand, academic self-efficacy primarily is individuals' confidence in their own learning abilities, emphasizing more on their self-assessment of competence and success. On the other hand, interface usability mainly related to the ease of operation and user experience on the MOOC platform, emphasizing the convenience and smoothness of user interaction (Cho et al., 2009). These two concepts are fundamentally different, as a result their relationship is not necessarily strong. Therefore, academic self-efficacy may not significantly moderate the impact of interface usability on the intention to continue using accounting MOOC. Furthermore, with the development of MOOC in China, there has been continuous improvement in the infrastructure of these platforms, which offers a much smoother and easier learning experience. The user experience has also gradually improved (Fang et al., 2019; Huang et al., 2012). University learners have already become accustomed to using MOOC platforms for learning, and the convenience of interface design may have already reached a relatively satisfactory level. Hence, learners are more concerned about their own learning abilities and feelings of success than the ease of platform operation. The moderating role of academic self-efficacy in interface usability may be relatively weak. Similarly, interface design aesthetics mainly focus on the MOOC platform's visual appeal and user experience, which has a weak correlation with academic self-efficacy and may not significantly moderate the impact of interface design aesthetics on the intention to continuous use. In the past decade of development, MOOC platforms have collaborated with higher education institutions to provide rich educational course resources and offer diverse learning content. These platforms have achieved a high level of interface design aesthetics, pursuing simple, intuitive, and visually appealing designs to provide a good user experience. Therefore, when using MOOC platforms, learners may not be concerned with the aesthetic factors of interface design but rather on the quality of course content and their understanding and mastery of the material. They pay more attention to the practicality of the courses and the reliability and usability of teaching resources. As a result, interface usability and interface design aesthetics are not the main concern for university learners, leading to this insignificant finding.

Table 3. Test results of structural model

| Hypothesis | Relationship | Path coefficient | T-values | P-values | Supported |
|---|---|---|---|---|---|
| H1 | IC->ATT | 0.49 | 4.317 | 0.000*** | YES |
| H2 | IDA-> ATT | 0.254 | 2.35 | 0.019* | YES |
| H3 | ATT->CI | 0.883 | 22.778 | 0.000*** | YES |
| H4 | PBC->CI | 0.177 | 11.798 | 0.000*** | YES |
| H5 | SN -> CI | 0.7 | 9.77 | 0.000*** | YES |
| H6a | EF*IC->ATT | 0.113 | 0.826 | 0.409 | NO |
| H6b | EF*IDA->ATT | -0.122 | 0.882 | 0.378 | NO |
| H6c | EF*ATT->CI | 0.063 | 2.857 | 0.004** | YES |
| H6d | EF*PBC->CI | 0.06 | 2.485 | 0.013* | YES |
| H6e | EF*SN->CI | 0.083 | 3.198 | 0.001* | YES |

*p-value < .05, **p-value < .01, ***p-value < .001.

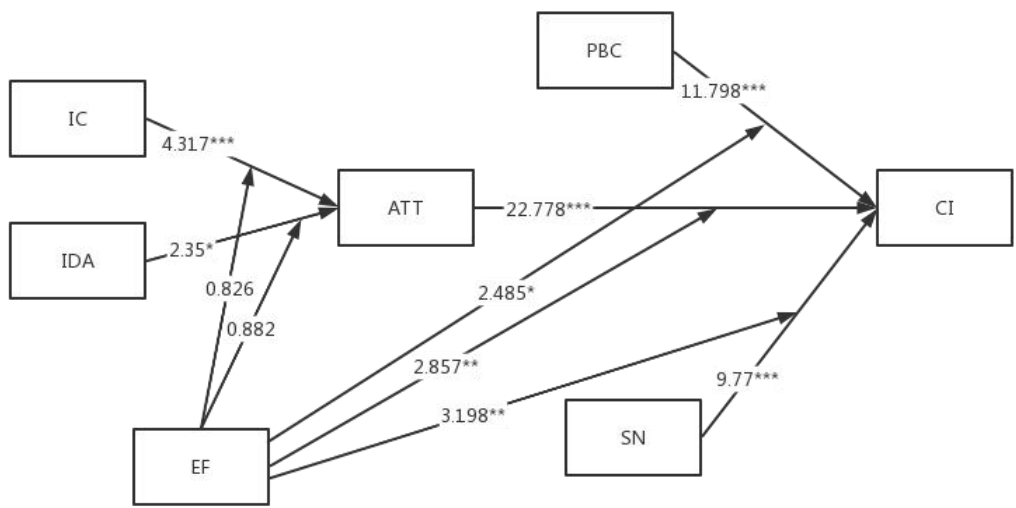

D. SHANG*a,b, Q. CHENa, X. GUO a, H. JINa, S. KE a, M. LI a

**Figure 2. Structural model test diagram**
(Note:IC,Interface convenience; IDA,Interface design aesthetics; ATT,attitude; PBC,Perceived behavioral control; SN,Subjective norm; EF, Academic self-efficacy; CI,continuance learning intention)

*p-value＜0.05, **p-value＜0.01，***p-value＜0.001

## 5. Discussions and conclusions
### 5.1 Theoretical implications
As information technology develops rapidly, digital education is on the rise. Massive open online courses (MOOC)

have become a popular trend of learning, providing convenient and flexible learning opportunities for learners. As an important professional course, accounting MOOC are of great significance to accounting learners. This study makes several contributions to the experience of continued usage intention in accounting MOOC and educational backgrounds. Firstly, based on the Theory of Planned Behavior, combined with a higher education background, an integrated theoretical framework is proposed, making a contribution to theoretical research. According to the analysis results, individuals intentions are often influenced by personal attitudes, perceived behavioral control, and subjective norms. This integrated theoretical framework comprehensively considers cognitive factors, emotional factors and environmental factors, which is beneficial for understanding and explaining the intention to continue using accounting MOOC in higher education contexts.

Moreover, previous studies have empirically investigated the aspects that influence the continuous intention to use accounting MOOC. These studies have examined learner characteristics (Jin & Shang, 2022; Rekha et al., 2023) as well as technological features (Shin & Song, 2022; Li et al., 2022; Rohan et al., 2021). However, these studies mainly focused on science and language courses (Apricio et al., 2019; Li et al., 2018; Psotka & Chen, 2019), with few empirical study in the area of accounting and other social science disciplines. To a large extent, the potential influencing factors between the accounting field in business and the previous professional education fields may differ (Fung, 1996; Tavani et al., 2016). Therefore, the research incorporates a new perspective on system functionality by investigating the influence of interface convenience (IC) and interface design aesthetics (IDA) concerning the intention to maintain the use of accounting MOOC. The results show that both interface convenience and interface design aesthetics significantly and directly contribute to the intention to continue using accounting MOOC. At the same time, interface convenience and interface design aesthetics also have a significant positive impact on the continued usage intentions of accounting MOOC through attitudes.

Furthermore, through empirical research, it has been found that the Theory of Planned Behavior, consisting of attitude, perceived behavioral control, and subjective norms, not only does it have a essential influence on the intention to persist in using MOOC but also on academic self-efficacy. Additionally, academic self-efficacy acts as a moderator between the Theory of Planned Behavior and the willingness to continue using accounting MOOC. As a result, the application of the Theory of Planned Behavior in accounting MOOC has been expanded. On one hand, this study extends the application scope of the Theory of Planned Behavior, initially employed to forecast consumer behavior and health behavior. This can broaden the application domain of the Theory of Planned Behavior and enrich its application in different fields. On the other hand, the boundary effect of academic self-efficacy is identified. It plays a moderating role between attitude, perceived behavioral control, subjective norms, and the intention to keep usage of accounting MOOC. This also expands the theoretical boundaries of academic self-efficacy. Therefore, the study results gain a deeper insight into the influencing factors behind the continued usage intention of accounting MOOC. In general, this study enriches existing literature, addresses shortcomings in the study of MOOC determinants and the social science field, and promotes the development of educational science research.

*5.2 Practical implications*

This study carries useful realistic suggestions for the operation of accounting MOOC. Through depth research on the motivation and behavior of MOOC, it can guide the optimization of MOOC platform operations, improve online learning effectiveness, and broaden learning channels, thereby promoting the development of accounting MOOC. Studying these factors can directly guide optimizations in content design, teaching methods, and interaction styles on MOOC platforms. By improving the interface convenience and interface aesthetic design of the accounting MOOC platform, learners will have a positive attitude towards its use (Shang & Wu, 2019).   As a large-scale online learning method, accounting MOOC have significant implications for learners' future career development. By studying the factors, individuals can receive guidance on motivation and behavior when participating in online learning. Furthermore, individuals can have the ability to self-assess and modify factors such as perceived behavioral control and subjective norms. This will allow them to increase their initiative and self-discipline in online learning, improve their accounting skills, and promote their professional development. At the same time, based on the characteristics and needs of different groups, MOOC platform offer personalized instruction, design tailored learning activities, and provide resource support due to individual needs, which expands learning channels, and meets personalized learning needs.

In addition, research results also indicate that academic self-efficacy significantly moderates the impacts of attitudes, perceived behavioral control, and subjective norms on the continuous intention to use accounting MOOC. Understanding the positive role of academic self-efficacy in the continuous intention to use MOOC can guide teachers to create more opportunities in accounting MOOC to help learners establish and strengthen their academic self-efficacy. By providing clear learning goals, appropriate challenges, teachers can help learners develop confidence, improve academic self-efficacy, and promote a continuous intention to use MOOC (Cho et al., 2009). Educators can gain insights into learners' motivations and psychological states by offering guidance, encouragement, and support throughout the learning process. This assistance helps learners to overcome challenges and setbacks, boosting their academic self-efficacy, and fostering a sustained intention to utilize accounting MOOC. As for MOOC designers, a thorough understanding of the moderating role of academic self-efficacy can help them better understand students' learning motivations and needs. So decision-makers can develop appropriate policies and measures, such as implementing learning achievement rewards, to encourage individuals and boost their belief and confidence in achieving positive learning outcomes. Not only does this encourage students to maintain their continuous intention in MOOC, but it also makes it easier for them to adopt it more widely.

*5.3 Conclusions, limitations and future directions*

This research is based on the Theory of Planned Behavior and explores the effect of emotional characteristics and technology-oriented factors on the intention to continue using MOOC in business accounting education. Through a questionnaire survey and structural equation model analysis of 176 university students, the study identified factors and relationships that affect the intention to continue using MOOC for accounting education. The research results have theoretical and practical implications, providing references for improving the user experience and enhancing the willingness to continue using MOOC. Furthermore, this study highlights the significance of academic self-efficacy in shaping the intention to persistently utilize MOOC. It serves as a crucial moderator, significantly influencing the impact of attitude, perceived behavioral control, and subjective norms on continuance learning intention. This research not only enriches the literature on MOOC and business disciplines but also enhances the understanding of the relationship between the Theory of Planned Behavior and the intention to continue using MOOC. Hence, this study builds upon existing literature to explore the influencing factors of intention to continue using MOOC, using accounting as an example, in the fields of educational theory and social sciences.

Due to the following reasons, discuss the limitations of the research is very essential: Firstly, the sample is restricted to 176 university students from a provincial university in China. Due to regional and cultural differences, the ability of the research findings may be restricted. Secondly, this study utilizes partial least squares structural equation modeling for analysis, which can only explore the relationships between variables and may have some biases. To gain a more comprehensive understanding, mixed methods, such as case studies or experimental designs, could be employed to explore different experiential and cognitive aspects. Furthermore, research could consider potential variations regarding learners of different gender, educational experience, MOOC learning experience, and majors to identify better learning strategies. To enhance the understanding and estimation of learners' behavioral intentions, cognition, and experiences in MOOC or other online learning environments, the application of mixed methods can be further applied. Additionally, the inclusion of cognitive neuroscience research should be considered, as it has increasingly influenced the development of theories in fields such as cognitive psychology and education. The Theory of Planned Behavior is a social psychology theory, and the use of cognitive psychology and neuroscience approaches can provide new insights through laboratory experiments and computational analyses, proposing new frameworks for future research. Therefore, it is recommended that future studies should consider these perspectives to better understand learners' behavioral intentions, cognition, ultimately yielding more persuasive results.

**Statements and Declarations**

Conflicts of interests: No financial or non-financial conflict of interest exits in the submission of this manuscript, and manuscript is approved by all authors for publication. We would like to declare that the work described was original research that has not been published previously, and not under consideration for publication elsewhere, in whole or in part.

## Appendix: Questionnaire measurement

**The independent variable**

**Academic self-efficacy（EF）**

1. Compared with other students in accounting class I expect to do well.
2. I'm certain I can understand the accounting and business ideas taught in this course.
3. I expect to do very well in accounting and business class.
4. Compared with others in accounting and business class, I think I'm a good student.
5. I am sure I can do an excellent job on the problems and tasks assigned for accounting class.

**Attitude（ATT）**

Using accounting mooc is a good idea.
I like using accounting mooc.
It is desirable to use accounting mooc.

**Perceived behavioral control（PBC）**

Using accounting mooc system was entirely within my control.
I had the resources, knowledge, and ability to use accounting mooc.
I would be able to use the accounting mooc system well for learning process.

**Subjective norm（SN）**

People important to me support my use of accounting mooc.
People who influence me think that I should use accounting mooc.
People whose opinions I value prefer that I should use accounting mooc.

**Interface convenience (IC)**

IC1 When I use the accounting mooc system, I think I can easily recognize where the needed information is located.
IC2 When I use the accounting mooc system, I think I can easily recognize where I could communicate with instructors.
IC3 I think that the screen design of the accounting mooc system is harmonious.

**Interface design aesthetics (IDA)**

IDA1 There is no inconvenience for using the accounting mooc system due to its display size.

IDA2 I am content with the color and of the accounting mooc system.

IDA3 I am satisfied with the module design of the accounting mooc system.

**The dependent variable**

**MOOC Continuance learning intention（CI）**

I will use the accounting mooc system on a regular basis in the future.

I will frequently use the accounting mooc system in the future.

I will strongly recommend that others use it.

**NOTES ON CONTRIBUTORS**

XX

D. SHANG*a,b, Q. CHENa, X. GUO a, H. JINa, S. KE a, M. LI a